\documentclass[a4paper, oneside, twocolumn, notitlepage, 10pt]{extarticle_ecoc}
\usepackage{ecoc}

\usepackage{graphicx}
\usepackage{tikz}
\usepackage{pgfplots}
\usepackage{mathtools}
\usepgfplotslibrary{colorbrewer}
\usepgfplotslibrary{groupplots}
\usetikzlibrary{shapes.geometric}
\usepackage{pgfplotstable}
\pgfplotsset{compat=1.17}
\usetikzlibrary{backgrounds}
\usepackage{wrapfig}  
\usepackage{amsfonts}
\usepackage{multirow}
\usetikzlibrary{arrows}
\usetikzlibrary{decorations.markings}
\usetikzlibrary{matrix}
\usepackage{xcolor}
\usepackage{colortbl}
\usepackage{threeparttable}
\usepackage{environ}
\usepackage{multicol}

\newcommand{\SNR}{\ensuremath{\mathrm{SNR}}}

\begin{document}
\selectlanguage{english}    


\title{Impact of launch power optimisation in hybrid-amplified links}%


\author{
    H.~Buglia\textsuperscript{(1)}, 
    E.~Sillekens\textsuperscript{(1)},
    L.~Galdino\textsuperscript{(2)},
    R.~I.~Killey\textsuperscript{(1)}, and P.~Bayvel\textsuperscript{(1)}
}

\maketitle                  


\begin{strip}
 \begin{author_descr}
 
   \textsuperscript{(1)} Optical Networks Group, UCL (University College London), UK. 
   \textcolor{blue}{\uline{henrique.buglia.20@ucl.ac.uk}}.

   \textsuperscript{(2)} Corning Optical Communications, Ewloe, CH5 3XD, United Kingdom.

 \end{author_descr}
\end{strip}

\setstretch{1.1}
\renewcommand\footnotemark{}
\renewcommand\footnoterule{}


\begin{strip}
  \begin{ecoc_abstract}
Per-channel launch power optimisation in a hybrid-amplified link with optimised pump powers and wavelengths is described. Compared to using the optimum spectrally uniform launch power, an average SNR gain of 0.13~dB is obtained against 0.56~dB for the same system operating with lumped amplifiers only. \textcopyright2024 The Author(s)
  \end{ecoc_abstract}
\end{strip}

\setlength{\abovedisplayskip}{3pt}
\setlength{\belowdisplayskip}{3pt}

\section{Introduction}
\vspace{-0.1cm}

To achieve the maximum throughput in ultra-wideband (UWB) systems using hybrid (Raman + rare-earth doped fibre) amplifiers, optimisation of launch power (LP), pump powers and wavelength allocation is required~\cite{buglia2024throughput,viacheslav}. Such optimisation has become possible thanks to closed-form expressions from the Gaussian noise model in the presence of Raman amplification~\cite{arxiv_modulation,buglia2023closedraman,buglia2024throughput}, which estimates the nonlinear interference (NLI) noise and amplified spontaneous emission (ASE) noise in real-time for UWB system operating with hybrid amplifiers.

LP optimisation for UWB has been extensively analysed for lumped-amplified links in~\cite{power1,power2,power2,power3,power4,power5,power6,power7,power8}. It was shown in~\cite{power2} that relevant gains in performance can be achieved by allocating higher power in the S-band and lower in the L-band to counteract the inter-channel stimulated Raman scattering (ISRS) effect. In~\cite{power2} it was also shown that these gains in performance depend on the system configuration including span lengths and optical bandwidths. However, for the case of hybrid-amplified links, very little research has been carried out to date. This is due to the very recent availability of real-time models~\cite{buglia2023closedraman,arxiv_modulation,buglia2024throughput} as well complex interactions between ISRS, pump powers and wavelengths, and the signal LP.

Indeed, the first theoretical UWB demonstrations of hybrid-link optimisation were reported in~\cite{buglia2024throughput,viacheslav} for backward pumped Raman systems. In~\cite{buglia2024throughput}, a spectrally uniform LP profile was used to maximise the system throughput, whereas in~\cite{viacheslav}, per-channel LP optimisation was performed. However, in the latter, no quantification of the gains obtained by optimising the LP was reported. This is important as LP optimisation is computationally complex and requires additional shaping components in the deployed systems, increasing the overall system cost and power consumption. 

In this work we quantify, for the first time, the gains in total throughout obtained by per-channel LP optimisation in hybrid-amplified links and compare the values with those obtained for lumped systems. A capacity-maximising hybrid amplifier is described, designed using a particle swarm optimisation (PSO) algorithm, where the launch power as well as forward and backward pumps are optimised in terms of power and wavelength to maximise system throughput. The results are compared to an optimum spectrally uniform LP profile. A similar analysis is carried out for a system with the same parameters but designed to operate using lumped amplifiers only.

\begin{figure}[t!]
\hspace*{-.6cm}
\vspace*{-.23cm}
    \begin{tikzpicture}[baseline]
\begin{axis}[
legend cell align=left,
title style={at={(axis cs:1540,0.239)}},
legend style={font=\footnotesize, at={(rel axis cs:0,1)}, anchor=north west},
width=8.0cm, height = 5.0cm,
xlabel={Wavelength [nm]},
ylabel={Attenuation [dB/km]},
grid=both,
xtick={1400,1430,1460,1490,1520,1550,1580,1610},
ymax=0.26,ymin=0.19,
xmin=1400,xmax=1610,
xticklabel style={/pgf/number format/1000 sep=},
    ytick distance=0.01,
    xtick distance=40,
]

\addplot[black,very thick] table[x=wav,y=att] {Data/ECOC_attenuation.txt};

\node[anchor=west] (source) at (axis cs:1430-6,0.255){E};
\node[anchor=west] (source) at (axis cs:1490-6,0.255){S};
\node[anchor=west] (source) at (axis cs:1540-6,0.195){C};
\node[anchor=west] (source) at (axis cs:1588-6,0.195){L};

 \draw[draw = green!60!black!80,>=stealth, ->,line width=1.2pt] (axis cs: 1455,0.2)--(axis cs:1455,0.215);
\draw[draw = green!60!black!80,>=stealth, ->,line width=1.2pt] (axis cs: 1410,0.2)--(axis cs:1410,0.215);
\draw[draw = green!60!black!80,>=stealth, ->,line width=1.2pt] (axis cs: 1405,0.2)--(axis cs:1405,0.215);
\node[above] at (axis cs: 1445,0.19) {Raman pumps};

 \draw[draw = blue!60!black!80,>=stealth, ->,line width=1.2pt] (axis cs: 1483,0.2)--(axis cs:1483,0.215);
\draw[draw = blue!60!black!80,>=stealth, ->,line width=1.2pt] (axis cs: 1452,0.2)--(axis cs:1452,0.215);
\draw[draw = blue!60!black!80,>=stealth, ->,line width=1.2pt] (axis cs: 1437,0.2)--(axis cs:1437,0.215);
\draw[draw = blue!60!black!80,>=stealth, ->,line width=1.2pt] (axis cs: 1428,0.2)--(axis cs:1428,0.215);
\draw[draw = blue!60!black!80,>=stealth, ->,line width=1.2pt] (axis cs: 1422,0.2)--(axis cs:1422,0.215);


\end{axis}

 \begin{axis}[tiny,anchor=north east,at={(rel axis cs:1,1)}, anchor=north east,
 legend cell align=left,
legend style={font=\footnotesize, at={(rel axis cs:1,1)}, anchor=north east},
xmin=0,xmax=25,
yticklabel style={
        /pgf/number format/fixed,
        /pgf/number format/precision=5
},
xtick = {0,5,10,15,20,25},
ylabel={Raman gain [1/km/W]},
x label style={at={(axis description cs:0.5,-0.1)},anchor=north},
y label style={at={(axis description cs:-0.35,0.48)},anchor=north},
xlabel={Frequency separation [THz]},
ymin=0,ymax=0.50,
ytick = 
{0,0.1,0.2,0.30,0.40,0.50},
ytick distance=0.025,
grid=both,
]
  \begin{scope}[on background layer]
    \draw[fill=white] (-9,-0.14) rectangle (26.5,0.53);
  \end{scope}   
               \addplot[very thick] table[x=freq,y= gain] {Data/ECOC_RamanProfile.txt};
 \end{axis}

\end{tikzpicture}
\caption{Fibre attenuation coefficient and Raman gain spectrum for the fibre used, and optimised forward (green arrows) and backward (blue arrows) pump wavelength allocation.}
\label{fig:attenuation}
\end{figure}
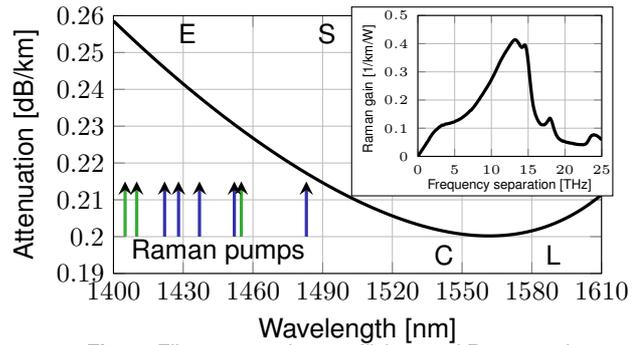

\vspace{-0.1cm}
\section{Transmission system setup}

It was assumed that the modelled system uses a hybrid amplification technology, consisting of two stages: a forward+backward-distributed Raman amplifier followed by an ideal EDFA. Transmission of $N_{\text{ch}}$=131 channels spaced by 100~GHz and centred at 1550~nm was modelled. Each channel was modulated at the symbol rate of 96~GBd, resulting in a total bandwidth of 13~THz~(105~nm), spanning 1500~nm to 1605~nm (transmission over the S-, C- and L- bands). Spectral gaps of 10~nm and 5~nm are assumed, respectively, between the S/C and C/L bands. The noise figure of each lumped amplifier placed at the end of each span is 7~dB, 4.5~dB, and 6~dB in the S-, C- and L- bands, respectively. Gaussian constellations, ideal lumped amplification, and ideal transceivers are assumed.

The signal transmission was evaluated over 10~$\times$~100~km spans (total distance of 1000~km). The  optical fibre  was assumed to have wavelength-dependent attenuation and the Raman gain profile compliant with Fig.~\ref{fig:attenuation}, and an effective area of $80~\mu\text{m}^2$, resulting in a nonlinear coefficient $\gamma = 1.16\text{ W}^{-1} \text{km}^{-1}$. 
Dispersion parameters considered are $D = 16.5~\text{ps }\text{nm}^{-1}\text{km}^{-1}$, $S = 0.09~\text{ps }\text{nm}^{-2}\text{km}^{-1}$. For the distributed Raman amplification, forward and backward pumps were placed in the E- and S-bands, as shown in Fig.~\ref{fig:attenuation}, and their wavelengths and powers, and the per-channel system LP were optimised to maximise the system throughput.

\vspace{-0.1cm}
\section{Launch power and pump optimisation}

\begin{figure}[t!]
\hspace*{-.6cm}
      \begin{tikzpicture}[baseline]
    \begin{axis}[
    unbounded coords=jump,view={50}{20}, grid=both,
    legend cell align=left,
legend style={font=\footnotesize, at={(rel axis cs:0.5,0.6,1.15)}, anchor=north},
width=7cm,
    x label style={rotate=-20},
     y label style={rotate=15},
    xlabel={Distance [km]},
ylabel={Wavelength [nm]},
zlabel={Power profile [dBm]},
x tick label style = {text width = 1.0cm, align = center, rotate = 70},
xtick={0,20,40,60,80,100},
ytick={1500,1550,1600},
zmin=-14,zmax=6,
ymin=1490,ymax=1610,
xmin=0,xmax=100,
xticklabel style={/pgf/number format/1000 sep=},
yticklabel style={/pgf/number format/1000 sep=},
  ytick distance=134,
  ztick distance=2,
    ]
    
      \addplot3[surf,color=red,opacity=0.4] file {Data/ECOC_powerprofile.txt};
      \addlegendentry{Forward + Backward Raman Amplification}
    \end{axis}
\end{tikzpicture}
\caption{Per-channel LP evolution with distance for the optimised launch and pump powers and pump wavelengths.}
\label{fig:power_3D}
\end{figure}
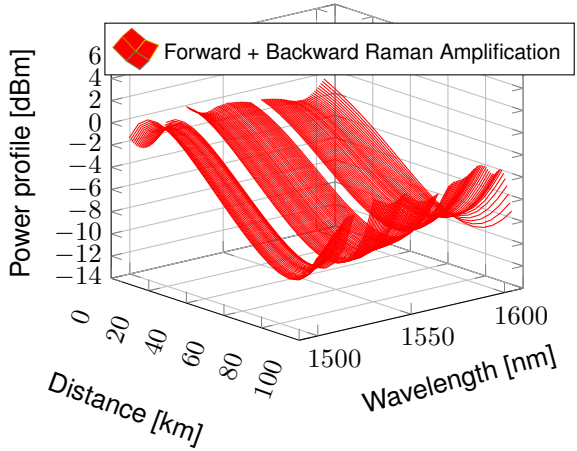

LP and pump were optimised over a single-span transmission based on a PSO algorithm from the Matlab global optimisation toolbox. The cost function to be maximised was the total system throughput, given by: $C = \sum_{i = 1}^{N_{\text{ch}}} 2\cdot\log_2(1+\SNR_i)$, where $\SNR_i$ is the SNR of the channel $i$, calculated using the model in~\cite{buglia2023closedraman,buglia2024throughput}.
Given the large number of degrees of freedom of this problem, the optimisation was divided into two steps. Firstly, pump powers and wavelengths were optimised with a spectrally uniform LP profile. Secondly, the pump powers and wavelengths found previously were kept constant and the channel LP was optimised.   

For the first step, 6 forward plus 6 backward pumps were placed in the E- and S-bands (1405~nm-1490~nm).  The PSO algorithm had 13 variables to be optimised (12 pumps + the total LP). The number of particles was chosen to be 50 with a maximum of 50 iterations selected as the stopping criterion. For the algorithm bounds, we allowed the total channel
LP to vary between 10~dBm and 25~dBm, and the power of each pump from
0~mW to 250~mW. The optimisation resulted in an optimum spectrally uniform LP of -1.74~dBm per channel, corresponding to a total LP of 18.75~dBm; 3 forward pumps with non-negligible power, with wavelengths located at 1405~nm, 1410~nm and 1455~nm, and powers of 153.6~mW, 240.1~mW and 31.3~mW, respectively; and 5 backward pumps with non-negligible power, with wavelengths located at 1422~nm, 1428~nm, 1437~nm, 1452~nm, and 1483~nm, and powers of 249.2~mW, 31.7~mW, 250~mW, 80~mW and 225.3~mW, respectively.

For the second step, the pump powers and wavelengths found previously were kept and the per-channel LP was optimised to maximise the system throughout. In this step, the PSO algorithm had $N_{\text{ch}}$ = 131 variables to be optimised, corresponding to the number of WDM channels.
The number of particles was chosen to be $10 \cdot N_{\text{ch}}$ with their values, i.e., the LP for each channel, ranging from -10~dBm to 10~dBm.
A maximum of 75 iterations was selected as the stopping criterion. The optimum per-channel LP evolution of the optimised hybrid amplifier with distance is shown in Fig.~\ref{fig:power_3D}. This optimisation was also carried out for a system operating with lumped amplification only. In that case, the per-channel launch-power bounds ranged from -5~dBm to 15~dBm, and the remaining parameters were kept the same. 
Finally, for both cases, the optimum per-channel LP obtained from the PSO algorithm was passed through a second-order Savitzky-Golay filter with a 7-channel window, resulting in a smooth per-channel LP that achieved a total throughput within 0.06~Tb/s of that obtained from the PSO.

\begin{figure}[t]
\vspace*{-.6cm}
\hspace*{-1cm}
\begin{tikzpicture}[baseline]
\begin{groupplot}[group style={group size=1 by 2, vertical sep=2cm,},width=8cm,height=4.5cm]
\nextgroupplot[
unbounded coords=jump,
title = \footnotesize (a) Hybrid amplification,
legend style={font=\footnotesize, at={(rel axis cs:1,0)}, 
legend columns=3,anchor=south east},
xlabel={Wavelength [nm]},
ylabel={Launch power [dBm]},
grid=both,
xtick={1490,1505,1520,1535,1550,1565,1580,1595,1610},
ymax=2.5,ymin=-5,
xmin=1490,xmax=1610,
xticklabel style={/pgf/number format/1000 sep=},
  ylabel near ticks,
    ytick distance=1.25,
    xtick distance=30,
]

\node[anchor=west] (source) at (axis cs:1505,1.9){S};
\node[anchor=west] (source) at (axis cs:1542,1.9){C};
\node[anchor=west] (source) at (axis cs:1582,1.9){L};







\addlegendentry{Optimum}
\addplot[Set1-B,thick] table[x=wav,y=optimum] {Data/ECOC_Power_hybrid.txt};
\addlegendentry{Spectrally uniform}
\addplot[Set1-C,thick,dashed] table[x=wav,y=uniform] {Data/ECOC_Power_hybrid.txt};


\nextgroupplot[
unbounded coords=jump,
title = \footnotesize (b) Lumped amplification,
legend style={font=\footnotesize, at={(rel axis cs:0,0)}, legend columns=2,anchor=south west},
xlabel={Wavelength [nm]},
ylabel={Launch power [dBm]},
grid=both,
xtick={1490,1505,1520,1535,1550,1565,1580,1595,1610},
ymax=7.5,ymin=0,
xmin=1490,xmax=1610,
xticklabel style={/pgf/number format/1000 sep=},
/pgf/number format/fixed,
  ylabel near ticks,
    ytick distance=1.25,
    xtick distance=15,
]

\node[anchor=west] (source) at (axis cs:1505,7){S};
\node[anchor=west] (source) at (axis cs:1542,7){C};
\node[anchor=west] (source) at (axis cs:1582,7){L};






\addlegendentry{Optimum}
\addplot[Set1-B,thick] table[x=wav,y=optimum] {Data/ECOC_Power_lumped.txt};
\addlegendentry{Spectrally uniform}
\addplot[Set1-C,thick,dashed] table[x=wav,y=uniform] {Data/ECOC_Power_lumped.txt};

\end{groupplot}

\end{tikzpicture}
\caption{Per-channel and spectrally uniform optimum LP profiles for (a) hybrid and (b) lumped amplification.}
\label{fig:power}
\end{figure}
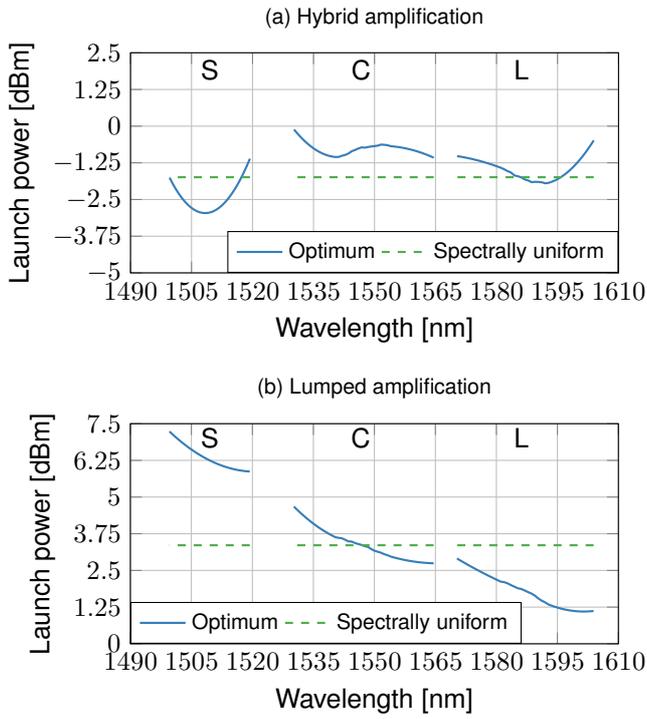

\begin{figure}[t]
\vspace*{-.6cm}
\begin{tikzpicture}[baseline]
\begin{groupplot}[group style={group size=1 by 2, vertical sep=2cm,},width=8cm,height=4.5cm]
\nextgroupplot[
clip=false,
legend entries={plot A},
 legend image code/.code={%
 \draw[dashed] (0cm,-0.1cm) -- (0.5cm,-0.1cm);
 \draw[solid]  (0cm, 0.1cm) -- (0.5cm, 0.1cm);
    },
unbounded coords=jump,
title = \footnotesize (a) Hybrid amplification,
legend style={font=\footnotesize, at={(rel axis cs:1,0)}, 
legend columns=3,anchor=south east},
xlabel={Wavelength [nm]},
ylabel={SNR [dB]},
grid=both,
xtick={1490,1505,1520,1535,1550,1565,1580,1595,1610},
ymax=30,ymin=16,
xmin=1490,xmax=1610,
xticklabel style={/pgf/number format/1000 sep=},
  ylabel near ticks,
    ytick distance=2,
    xtick distance=15,
]

\node[anchor=west] (source) at (axis cs:1505,29){S};
\node[anchor=west] (source) at (axis cs:1542,29){C};
\node[anchor=west] (source) at (axis cs:1582,29){L};


\addlegendentry{ASE}
\addplot[Set1-D,thick] table[x=wav,y=ASE] {Data/ECOC_SNR_hybrid_optimum.txt};
\addplot[Set1-D,thick,dashed,forget plot] table[x=wav,y=ASE] {Data/ECOC_SNR_hybrid_uniform.txt};
\addlegendentry{NLI}
\addplot[Set1-E,thick] table[x=wav,y=NLI] {Data/ECOC_SNR_hybrid_optimum.txt};
\addplot[Set1-E,thick,dashed,forget plot] table[x=wav,y=NLI] {Data/ECOC_SNR_hybrid_uniform.txt};
\addlegendentry{Total}
\addplot[Set1-A,thick] table[x=wav,y=Total] {Data/ECOC_SNR_hybrid_optimum.txt};
\addplot[Set1-A,thick,dashed,forget plot] table[x=wav,y=Total] {Data/ECOC_SNR_hybrid_uniform.txt};


\nextgroupplot[
legend entries={plot A},
 legend image code/.code={%
 \draw[dashed] (0cm,-0.1cm) -- (0.5cm,-0.1cm);
 \draw[solid]  (0cm, 0.1cm) -- (0.5cm, 0.1cm);
    },
unbounded coords=jump,
title = \footnotesize (b) Lumped amplification,
legend style={font=\footnotesize, at={(rel axis cs:1,0)}, legend columns=3,anchor=south east},
xlabel={Wavelength [nm]},
ylabel={SNR [dB]},
grid=both,
xtick={1490,1505,1520,1535,1550,1565,1580,1595,1610},
ymax=26,ymin=8,
xmin=1490,xmax=1610,
xticklabel style={/pgf/number format/1000 sep=},
/pgf/number format/fixed,
  ylabel near ticks,
    ytick distance=2,
    xtick distance=15,
]

\node[anchor=west] (source) at (axis cs:1505,25){S};
\node[anchor=west] (source) at (axis cs:1542,25){C};
\node[anchor=west] (source) at (axis cs:1582,25){L};

\addlegendentry{ASE}
\addplot[Set1-D,thick] table[x=wav,y=ASE] {Data/ECOC_SNR_lumped_optimum.txt};
\addplot[Set1-D,thick,dashed,forget plot] table[x=wav,y=ASE] {Data/ECOC_SNR_lumped_uniform.txt};
\addlegendentry{NLI}
\addplot[Set1-E,thick] table[x=wav,y=NLI] {Data/ECOC_SNR_lumped_optimum.txt};
\addplot[Set1-E,thick,dashed,forget plot] table[x=wav,y=NLI] {Data/ECOC_SNR_lumped_uniform.txt};
\addlegendentry{Total}
\addplot[Set1-A,thick] table[x=wav,y=Total] {Data/ECOC_SNR_lumped_optimum.txt};
\addplot[Set1-A,thick,dashed,forget plot] table[x=wav,y=Total] {Data/ECOC_SNR_lumped_uniform.txt};

\end{groupplot}

\end{tikzpicture}
\caption{SNR contributions after 10 spans using per-channel (continuous lines) and spectrally uniform (dashed lines) optimum LP profile for (a) hybrid and (b) lumped amplification.}
\label{fig:SNR}
\end{figure}
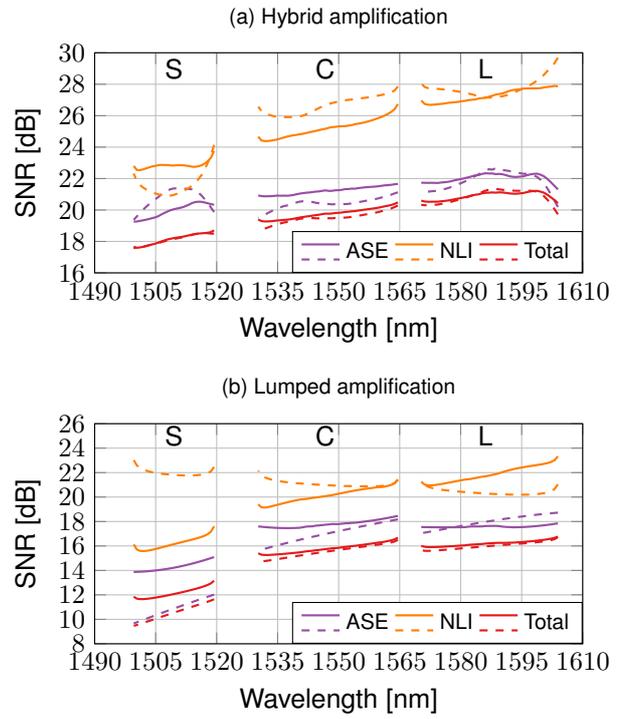

\vspace{-0.1cm}
\section{Results and Discussion}

The optimum per-channel and spectrally uniform LP profiles for the optimised hybrid amplifier are shown in Fig~\ref{fig:power} after one span. For the optimum per-channel LP, a total power of 19.14~dBm was obtained with a total per-channel variation of 2.85~dB. For the optimum spectrally uniform LP, the total power was 18.75~dBm. The same analysis was carried out and is shown in Fig.~\ref{fig:power}.b for an optimised system operating with lumped amplifiers only; in that case, these values were 24.44~dBm, with a total per-channel variation of 6.14~dB, and 23.85~dBm respectively.

Fig~\ref{fig:SNR} shows the system performance after 10 spans for the LP profiles shown in Fig.~\ref{fig:power}.a. For the optimised hybrid amplifier, using the optimum spectrally uniform LP, yields increased values of NLI noise and reduced values of ASE noise in the S-band, resulting mostly from the noise from forward Raman pumps within that band. On the other hand, reduced values of NLI noise and increased values of ASE noise are observed in the C- and L- bands which are most impacted by the noise from the backward Raman pumps. To counteract these noises,  the per-channel optimum LP allocates reduced power levels in the S-band and increased power levels in the C- and L-bands (see Fig.~\ref{fig:power}.a), increasing the system performance. 
The results of the same analysis carried out for the optimised lumped amplified system are shown in Fig~\ref{fig:SNR}.b. As is well known in such cases, the optimum per-channel LP allocates more power in the S-band to counteract the ISRS effect (see Fig.~\ref{fig:power}.b).

In terms of performance for the optimised hybrid amplifier (Fig.~\ref{fig:SNR}.a), using the per-channel optimum LP profile, yields an average total SNR of 19.78~dB, and total throughputs of 213.14~Tb/s and 141.64~Tb/s after 1 and 10 spans respectively. On the other hand, using the optimum spectrally uniform LP profile yields an average total SNR of 19.65~dB, and a total throughput of 212.48~Tb/s and 140.70~Tb/s, over 1 and 10 spans respectively. The same analysis for the optimised lumped amplifier (Fig.~\ref{fig:SNR}.b), yields an average total SNR value of 15.07~dB and total throughputs of 179.88~Tb/s and 108.66~Tb/s over 1 and 10 spans for the optimum per-channel LP profile. These values using a spectrally uniform LP profile are 14.50~dB, 175.72~Tb/s and 104.86~Tb/s respectively. 

In terms of gains obtained through the use of per-channel LP optimisation, compared to the spectrally uniform case, the hybrid-amplified system yields 0.13~dB in total SNR gain, and 0.66~Tb/s (0.3~\%) and 0.94~Tb/s (0.7~\%) throughput gains over 1 and 10 spans. For the lumped-amplified system, the gains with per-channel LP optimisation are significantly larger, at 0.56~dB, 4.16~Tb/s (2.4~\%) and 3.80~Tb/s (3.6~\%) respectively.

\vspace{-0.16cm}
\section{Conclusions}

The first analysis quantifying the gains of per-channel launch power optimisation in hybrid-amplified links was presented, where optimum spectrally uniform power profile was used for comparison. The modelled system used a hybrid amplifier with forward+backward Raman pumps, where launch power, pump powers and wavelengths were optimised to maximise the total throughput.

For hybrid-amplified systems, in comparison with lumped-amplified systems, it was found that pump power and wavelength optimisation can compensate for almost all the power fluctuations induced by the combination of Raman, ISRS and wavelength-depend fibre parameters, reducing the further gain in SNR achievable with per-channel launch power optimisation. For the optimised hybrid-amplified system only 0.13~dB SNR gain was due to the per-channel power optimisation. In contrast, per-channel power optimisation in the system with lumped amplification was responsible for a larger SNR gain of 0.56~dB.


\section{Acknowledgements}
This work was supported by EPSRC grants EP/R035342/1 TRANSNET
, EP/W015714/1 EWOC
, EPSRC studentship
(EP/T517793/1), and Microsoft Scholarship under the 'Optics for the Cloud' programme.

\printbibliography

\vspace{-4mm}

\end{document}